\documentclass[conference]{IEEEtran}
\IEEEoverridecommandlockouts
\usepackage{algorithm}
\usepackage{algpseudocode}
\usepackage{amsmath}
\usepackage{multirow} 
\usepackage{tabularx}
\usepackage{booktabs}
\usepackage{cite}
\usepackage{amsmath,amssymb,amsfonts}
\usepackage{graphicx}
\usepackage{textcomp}
\usepackage{xcolor}
\usepackage{subcaption}

\usepackage[inkscapelatex=false]{svg}
\usepackage[marginal]{footmisc}

\usepackage{authblk}
\def\BibTeX{{\rm B\kern-.05em{\sc i\kern-.025em b}\kern-.08em
    T\kern-.1667em\lower.7ex\hbox{E}\kern-.125emX}}
\begin{document}


\title{



EEG-ReMinD: Enhancing Neurodegenerative\\EEG Decoding through Self-Supervised State Reconstruction-Primed Riemannian Dynamics
}

\author{
Zirui Wang\textsuperscript{1}, Zhenxi Song\textsuperscript{1}$^{*}$, Yi Guo\textsuperscript{2}, Yuxin Liu\textsuperscript{1}, Guoyang Xu\textsuperscript{1}, Min Zhang\textsuperscript{1}, and Zhiguo Zhang\textsuperscript{1} \\
\textsuperscript{1}Harbin Institute of Technology, Shenzhen, China \\
\textsuperscript{2}Department of Neurology, Shenzhen People's Hospital, Shenzhen, China \\
\vspace{-1.5em} 
\thanks{ 
 \noindent\rule{0.5\linewidth}{0.25pt} 
 \\ 
$^{*}$Corresponding author: Zhenxi Song (songzhenxi@hit.edu.cn). This work is supported by the National Natural Science Foundation of China (Grant No. 62306089), the Shenzhen Science and Technology Program (Grant No. RCBS20231211090800003).}
}   
\maketitle

\begin{abstract}
The development of EEG decoding algorithms confronts challenges such as data sparsity, subject variability, and the need for precise annotations, all of which are vital for advancing brain-computer interfaces and enhancing the diagnosis of diseases. To address these issues, we propose a novel two-stage approach named Self-Supervised State Reconstruction-Primed Riemannian Dynamics (\textit{EEG-ReMinD}), which mitigates reliance on supervised learning and integrates inherent geometric features. This approach efficiently handles EEG data corruptions and reduces the dependency on labels.
\textit{EEG-ReMinD} utilizes self-supervised and geometric learning techniques, along with an attention mechanism, to analyze the temporal dynamics of EEG features within the framework of Riemannian geometry, referred to as Riemannian dynamics.
Comparative analyses on both intact and corrupted datasets from two different neurodegenerative disorders underscore the enhanced performance of \textit{EEG-ReMinD}.
\end{abstract}


\section*{I. Introduction}

Developing robust, generalizable, and interpretable EEG decoding algorithms is crucial for enhancing brain-computer interfaces and improving disease diagnosis. Neurodegenerative diseases such as Parkinson's Disease (PD) and Mild Cognitive Impairment (MCI) exemplify the challenges in this area, characterized by the subtle and varied manifestations in resting-state EEG data\cite{1,2,3,4}. While EEG holds significant potential for diagnostic applications\cite{5,6,9,10}, the variability between subjects and data sparsity complicates the development of effective models. These challenges underscore the need to explore suitable low-dimensional feature spaces for robust representation and to develop self-supervised learning (SSL) strategies that reduce reliance on clinical labels.

Given the decline in synchronous neural activity in neurodegenerative diseases, we explore EEG's low-dimensional space through graph mapping and geometric learning to develop robust representations. Research in Euclidean space utilizes Graph Neural Networks (GNN) to map spatial relationships between EEG channels, effectively capturing structural features through graph constructions \cite{11,12}. Transitioning to Riemannian space, the adaptations leverage the manifold's capacity to adapt to EEG’s inherent non-Euclidean structure \cite{14}. Utilizing the distinct geometric properties of Riemannian manifolds, these models harness invariant metrics to enhance robustness against the inherent non-stationarity and noise in EEG signals, thereby improving generalization across complex EEG data \cite{13,15,16,17,18}. Recent innovations in this domain include the deployment of a manifold-Euclidean combined model that specifically addresses the representation of spatio-temporal features of EEG data \cite{19}, and the employment of attention mechanisms to analyze sequences of functional brain networks in Riemannian space \cite{20}. Consequently, these advancements inspire us to integrate functional networks with geometric representation learning to enhance model interpretability.


However, these methods rely on supervised learning, which requires extensive labeled data. This is particularly challenging for neurodegenerative diseases, where subtle and variable EEG characteristics make accurate labeling difficult. Additionally, EEG data are prone to corruption, such as continuous segment and channel disruptions. Thus, developing self-supervised learning strategies is crucial to reduce label dependency and enhance model robustness by learning intrinsic representations that effectively handle EEG data corruptions. Reconstruction learning captures temporal and frequency features \cite{frame,frame2}. Contrastive learning enhances feature extraction by maximizing similarity between different samples \cite{contrast,contrast2}.



Despite these advances, current approaches have not fully leveraged graph construction and geometric representation learning to develop self-supervised algorithms. Therefore, this paper introduces a novel two-stage EEG decoding approach, termed \textit{Self Supervised State \textbf{Re}construction-Primed Rie\textbf{M}nn\textbf{i}a\textbf{n} \textbf{D}ynamics (EEG-ReMinD)} to overcome existing limitations. The main contributions of this paper can be summarized as follows:
1) We developed a two-stage EEG-ReMinD training strategy that primes robust representations via SSL, validated on MCI and PD datasets.
2) Our SSL framework is based on model internal state reconstruction, uniquely identifying states within time-varying geometric maps in Riemannian manifold spaces.
3) We facilitate state reconstruction from a Riemannian dynamics perspective by incorporating spatio-temporal filters, learnable positional encodings, and attention-based analysis of time-varying geometric maps.

\begin{figure*}[t]
    \centering
    \centering
    \includegraphics[width=0.8\textwidth]{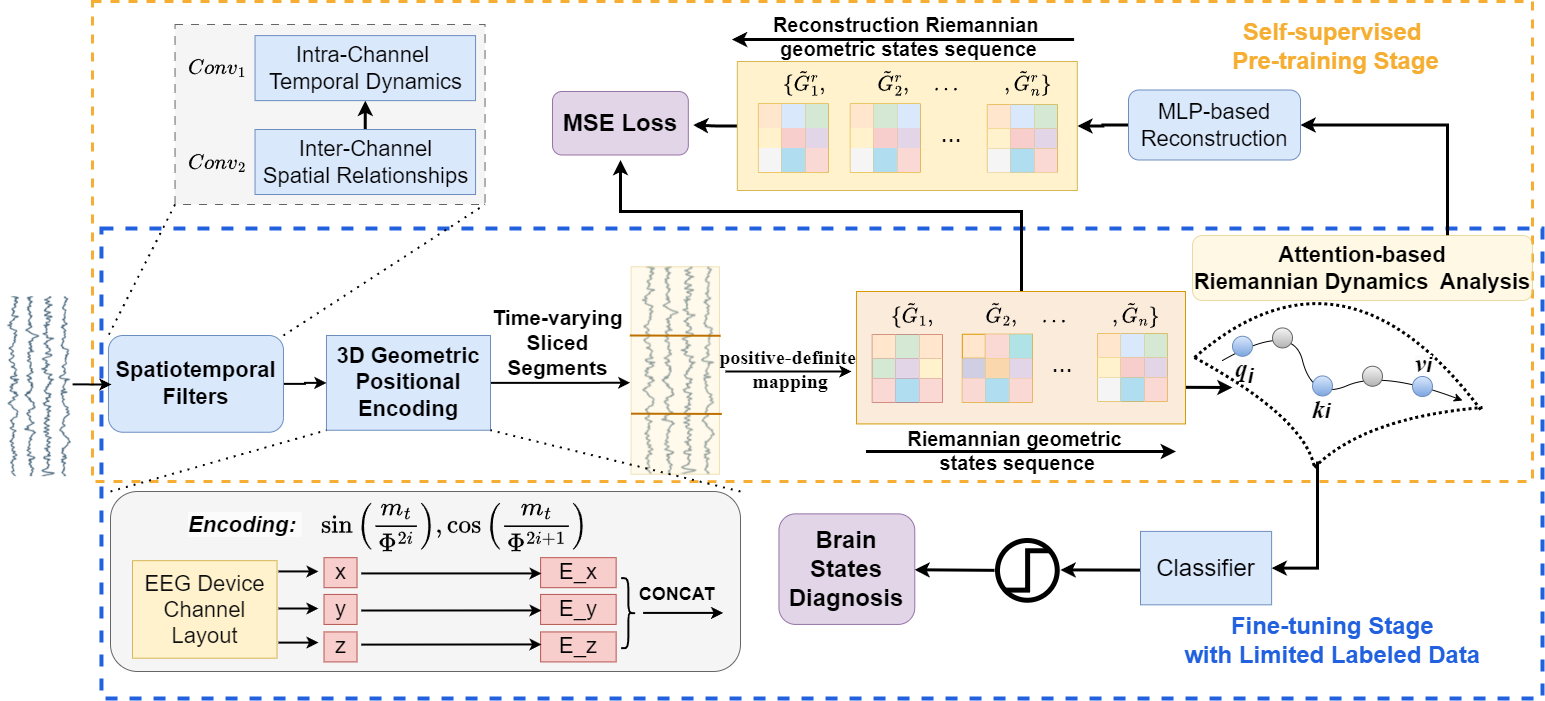} 
    \caption{Illustration of two-stage framework.}
    \label{model}
\end{figure*}

\section*{II. METHODS}

The proposed EEG-ReMinD model is illustrated in Fig.\ref{model} and consists of an unsupervised pre-training stage followed by a fine-tuning stage that utilizes limited labeled data.
The initial stage of the framework employs SSL devoid of labels, utilizing Riemannian manifold geometric representations. This phase comprises the systematic construction and subsequent reconstruction of Riemannian geometric features, which serve as the internal state within the trainable model (refer to Sections A and B). The second stage capitalizes on the pre-trained model to facilitate brain state recognition, utilizing only a sparse amount of labeled data (refer to Section C).
 
\subsection*{A. Construction of Riemannian Geometric States}

Let the multi-channel EEG signals be denoted by a matrix $X \in \mathbb{R}^{C \times T}$, where $C$ represents the number of channels and $T$ corresponds to the number of time points. 
To capture inter-channel relationships over time and intra-channel temporal dynamics, we utilized spatiotemporal filters comprising two convolutional layers to extract generalized representations across multiple channels and periods, enhancing robustness against noise and artifacts. The output of these filters yields a feature matrix, denoted as $\tilde{X}\in\mathbb{R}^{C\times T}$. This enables the subsequent construction of the Riemannian geometric state, effectively incorporating local channel correlations and temporal factors.

Traditional positional encoding focuses on incorporating sequence order information. Nevertheless, considering that the subsequent Riemannian geometric states should effectively capture the geometric relationships between electrodes, we introduce 3D-Geometric position encoding to integrate geometric positional information:
\begin{align}
    \small
    \text{Enc}(m_s, i) =
    \begin{cases} 
    \sin\left(\frac{m_s}{\Phi_i}\right), & i = 2k, \, k \in \mathbb{N}, \\
    \cos\left(\frac{m_s}{\Phi_i}\right), & i = 2k+1, \, k \in \mathbb{N}.
    \end{cases}
\end{align}

\begin{align}
    \small
    \text{GPE}(x_s, y_s, z_s, i) = \text{cat}\left[\text{Enc}(x_s, i), \text{Enc}(y_s, i), \text{Enc}(z_s, i)\right].
\end{align}

Here, $(x_s, y_s, z_s)$ represents the 3D coordinates of the electrodes, where $s$ denotes the $s$-th sensor. The variable $i$ represents the index of the feature dimension, and $\Phi$ is a constant. The final encoding $GPE$, is generated by concatenating the encodings of $x_t$, $y_t$, and $z_t$.
By incorporating this learnable 3-D geometric position encoding of the electrodes into the EEG features extracted by the CNN, the subsequent construction of the Riemannian geometric states can gain accurate positional information, thereby enhancing its robustness.

After incorporating 3-D geometric position encoding, the data is segmented into \(n\) sequence segments, denoted as \( \left\{ \tilde{X}_{1}, \tilde{X}_{2}, \dots, \tilde{X}_{n-1}, \tilde{X}_{n} \right\} \). For each of these segments, considering the correlation characteristics of the EEG signal between electrodes, we constructed the internal geometric states sequence (functional brain networks) to capture the functional connectivity between electrodes. As a result, the segments \( \left\{ \tilde{X}_{1}, \tilde{X}_{2}, \dots, \tilde{X}_{n-1}, \tilde{X}_{n} \right\} \) are transformed into  the internal geometric states sequence \( \left\{ G_{1}, G_{2}, \dots, G_{n-1}, G_{n} \right\} \), where each internal geometric state $G_{m_{i,j}}$ is defined as follows:
\begin{align}
G_{m_{i,j}} = \frac{\text{xcovar}(\tilde{X}_{m_{i}},\tilde{X}_{m_{j}})}{||\tilde{X}_{m_{i}}||||\tilde{X}_{m_{j}}||}, \quad 1 \le i,j \le C
\end{align}
where xcovar(·,·) denotes covariance function, $m$ presents the $m$th sequence. $X_{m_{i}}$ presents the  $m$th sequence's $i$th channel. $G_{m_{i,j}} $presents the  $m$th sequence's $i$th row  $j$th column value.

The internal geometric states sequence \( \left\{ G_{1}, G_{2}, \dots, G_{n} \right\} \) is projected from Euclidean space onto a Riemannian manifold via positive-definite mapping, forming the Riemannian geometric states sequence \( \left\{ \tilde{G}_{1}, \tilde{G}_{2}, \dots, \tilde{G}_{n-1}, \tilde{G}_{n} \right\} \). 

\subsection*{B. Reconstruction of Internal States}
To capture the temporal dependencies within the Riemannian geometric states sequence, we introduce an attention-based computation strategy based on Riemannian manifold, where $QKV$ and the corresponding attention computations are performed within the framework of Riemannian Dynamics.  
In this framework, we focus on time-varying behavior and evolution of the EEG data's Riemannian geometric representation, using Riemannian metrics to compute geometric properties such as distances in a non-Euclidean space. The advantage of using Riemannian Dynamics lies in their metric invariance properties, which enable the model to generalize well to complex EEG signals while also being robust to the non-stationarity and noise inherent in EEG signals.

We calculate the $Q,K,V$ of the brain network through bilinear mapping, as follows:
\begin{align}
Q_i  = W_q  \tilde{G}_{i} W_q^T, \quad K_i  = W_k  \tilde{G}_{i} W_k^T, \quad V_i = W_v  \tilde{G}_{i}W_v^T
\end{align}
Here, \( W_q \), \( W_k \), and \( W_v \) are the weight matrices of the linear mappings. Through the above bilinear mapping, $Q,K,V$ are constrained to be Symmetric Positive Definite(SPD) matrices on the manifold. 
The standard attention mechanism based on Euclidean distance cannot directly compute similarity in Riemannian space. To address this, we utilize the Log-Euclidean metric method\cite{15} which effectively calculates the center on the SPD manifold and uses Log-Euclidean distance as a similarity measure in Riemannian space, the geodesic distance from $Q_i$ to $K_j$ is given by:
\begin{align}
\text{LE-distance}(Q_i, K_j) = \|\text{Log}(Q_i) - \text{Log}(K_j)\|_F
\end{align}
The similarity calculation is as follows:
\begin{align}
\text S_{ij} = {Sim}(Q_i, K_j) = \frac{1}{1 + \log(1 + \text{LE-distance}(Q_i, K_j))}
\end{align}
Then the sub-matrix is normalized with the Softmax function to perform normalization along each row:
\begin{align}
    \small
    S' = \text{Softmax}(S) = \text{Softmax}\left( \left[S_{ij}\right]_{n \times n} \right) = \left[ S_{ij}' \right]_{n \times n}
\end{align}
where \( S_{ij}' = \frac{\exp(S_{ij})}{\sum_{k=1}^{m} \exp(S_{ik})}, \ \forall \ i,j \in \{1, \dots, n\} \).
Furthermore, The weighted Log-Euclidean mean\cite{mean} can be defined as:
\begin{align}
G(w_1, \dots, w_n, P_1, \dots, P_n) = \exp\left(\sum_{i=1}^{n} w_i \log(P_i)\right)
 \end{align}
where the weight of each SPD matrice$\{X_i\}_{l=1}^{n}$ is defined as $\{w_l\}_{l=1}^{n}$, $\{w_l\}_{l=1}^{n}$ satisfies the convexity constraint definition.

By utilizing the weighted Log-Euclidean mean,we combine \( V_1, \dots, V_k \) and the attention-score matrix to get the final the Riemannian geometric representation sequence $\{\tilde{V}_1, \tilde{V}_2, \dots, \tilde{V}_m\}$:
\begin{align}
\small
 \tilde{V}_i =  G( S'_{i1}, \dots,  S'_{in}, V_1, \dots, V_n) = \exp\left(\sum_{j=1}^{n} S'_{ij} \log(V_j)\right)
  \end{align}

To ensure that the Riemannian geometric representation effectively captures temporal and structural relationships as well as EEG time-varying behavior, we use a simple MLP layer to reconstruct the Riemannian geometric states sequence, \( \left\{ \tilde{G}_{1}, \tilde{G}_{2}, \dots, \tilde{G}_{n-1}, \tilde{G}_{n} \right\} \).
The reconstruction yields the sequence $ \{ \tilde{G}^{(r)}_1, \tilde{G}^{(r)}_2, \dots, \tilde{G}^{(r)}_{n-1}, \tilde{G}^{(r)}_n \}$ and then optimize the model by minimizing the Mean Squared Error (MSE) loss between the reconstructed sequence and the target sequence.
\begin{equation}
\{ \tilde{G}^{(r)}_1, \tilde{G}^{(r)}_2, \dots, \tilde{G}^{(r)}_{n-1}, \tilde{G}^{(r)}_n \} = \text{MLP}( \tilde{V}_1, \tilde{V}_2, \dots, \tilde{V}_n)
\end{equation}
\begin{equation}
\text{MSE} = \frac{1}{n} \sum_{i=1}^{n} \left\| \tilde{G}^{(r)}_i - \hat{G}_{i} \right\|^2 
\end{equation}
By reconstructing Riemannian geometric states sequence, the model retains the core structure of the EEG data at multiple levels, allowing it to recover crucial information even in the presence of poor data quality. This enhances the model's robustness to incomplete data, enabling it to remain resilient in the face of noise or anomalies in the input data. 
\begin{table*}[htbp]
\caption{Performance comparison with related methods on PD and MCI datasets based on N-fold cross-validation}
\centering
\begin{tabularx}{\textwidth}{X|XX|XX|cc}
\hline
\multirow{2}{*}{Methods} & \multicolumn{2}{c|}{PD (3-Fold)} & \multicolumn{2}{c|}{MCI (4-Fold)}  & \multicolumn{2}{c}{Comprehensive 
Results}\\
\cline{2-7}
 & Accuracy & F1 & Accuracy & F1 & Accuracy & F1 \\
\hline
Correlation\cite{21} & 66.67 (0.21) & 61.39 (0.78) & 65.34 (1.64) & 67.56 (2.58) & 66.01 & 63.99 \\
SPD Features\cite{22}  & 79.63 (0.27) & 79.46 (0.17) & 64.43 (0.77) & 68.18 (0.99) &72.03 &74.80 \\
Tensor-CSPNet\cite{23}                & 75.92 (6.41) & 73.20 (8.16) & 80.78 (7.93) & 77.70 (11.63) &78.35 &75.45 \\
MAtt\cite{19}                        & 79.63 (11.56) & 83.07 (7.56) & 81.97 (3.96) & 79.56 (8.29) &80.80 &81.32 \\
COMET\cite{24}                        & 75.47 (9.75) & 74.92 (14.17) & 73.25 (26.02) & 70.09 (24.82)&74.36 &72.51 \\
CTW \cite{25}                       & 75.93 (3.21) & 78.73 (2.20) & 75.40 (10.28) & 76.75 (9.80)&75.67 &77.74 \\
BNMTrans\cite{20}                     & 85.19 (8.49) & 83.25 (11.37) & 88.74 (4.61) & 88.18 (7.23)&86.97 & 85.72\\
EEG-ReMinD             & \textbf{88.89 (9.62)} & \textbf{88.85 (9.69)} & \textbf{91.06 (3.01)} & \textbf{91.07 (3.26)} & \textbf{89.97} & \textbf{89.96} \\
\hline
\end{tabularx}
\label{SOTA}
\end{table*}

\begin{table*}[htbp]
\caption{Performance comparison under different EEG data corruption types on both MCI and PD datasets}
\centering
\begin{tabularx}{\textwidth}{XX|XX|XX}
\hline
\multirow{2}{*}{Corruption} & \multirow{2}{*}{Method} & \multicolumn{2}{c|}{PD (3-Fold)}& \multicolumn{2}{c}{MCI (4-Fold)} \\ 
\cline{3-6}
 & & Accuracy & F1 & Accuracy & F1 \\ 
\hline
\multirow{3}{*}{Random Corruption} & EEG-ReMinD & \textbf{85.18 (8.49)} & \textbf{83.25 (11.37)}  & \textbf{84.23 (2.81) } & \textbf{82.56 (8.89) } \\
 & BNMTrans\cite{20}    & 83.33 (5.56) & 81.04 (8.49) & 78.61 (6.96) & 76.18 (9.59) \\
  & COMET\cite{24}   & 50.25 (3.92) & 33.40 (1.73)  & 55.66 (10.78)  & 46.72 (5.22) \\ 
\multirow{3}{*}{Segment Corruption} & EEG-ReMinD & \textbf{83.33 (5.56)}& \textbf{83.07 (7.56)}  & \textbf{81.96 (6.57)} & \textbf{81.53 (9.09)}\\
 & BNMTrans\cite{20}    & 79.62 (6.41) & 77.74 (9.64) & 77.47 (5.74)& 78.56 (8.79) \\
   & COMET\cite{24}   & 69.68 (0.45) & 68.98 (0.04)   & 60.23 (4.53) & 56.61 (2.78) \\ 
\multirow{3}{*}{Channel Corruption} &EEG-ReMinD & \textbf{85.18 (8.49)} & \textbf{83.25 (11.37)} & \textbf{80.78 (10.21)} & \textbf{80.15 (13.59)}\\
 & BNMTrans\cite{20}    & 81.48 (8.48) & 82.07 (7.22) & 76.43 (4.07) & 77.51 (5.22) \\
   & COMET\cite{24}   & 74.34 (1.02) & 73.98 (1.21) & 59.50 (7.09) & 49.17 (3.53)\\ 
\hline
\end{tabularx}
\label{corruption}
\end{table*}

\begin{table}[htbp]
\caption{Ablation study of EEG-ReMinD}
\centering
\begin{tabularx}{\columnwidth}{X|XX|XX}
\hline
\multirow{2}{*}{Strategies} & \multicolumn{2}{c|}{PD (3-Fold)} & \multicolumn{2}{c}{MCI (4-Fold)} \\
\cline{2-5}
 & Accuracy & F1 & Accuracy & F1 \\
\hline
ReMinD & \textbf{88.89 (9.62)} & \textbf{88.85 (9.69)} & \textbf{91.06 (3.01)} & \textbf{91.07 (3.26)} \\
ManiSSL & 85.19 (9.48) & 83.46 (11.67) & 84.33 (5.49) & 83.92 (5.78) \\
EucliSSL & 81.48 (8.49) & 77.75 (14.13) & 80.93 (4.12) & 81.43 (4.31) \\
w/o SSL & 83.33 (5.56) & 81.89 (9.50) & 86.41 (8.37) & 85.02 (11.07) \\
w/o 3D-Pos & 79.62 (11.56) & 74.37 (21.23) & 79.74 (9.53) & 81.40 (10.00) \\
w/o Filter & 74.07 (3.21) & 70.71 (9.23) & 77.52 (2.43) & 75.79 (4.79) \\
\hline
\end{tabularx}
\label{Ablation}
\end{table}

 \subsection*{C. Limited Labeled Data Fine-tuning Stage}
 By minimizing the reconstruction loss of Riemannian geometric representations during the pre-training phase, we develop a Riemannian geometric encoder with enhanced representative capabilities.
 Then, the pre-trained encoder can process the new EEG data and generate the Riemannian representations for classification. This stage requires only 10\% labels of training data for supervised fine-tuning, and tests on the remaining fold to obtain the cross-validation results.

\section*{ III.  MATERIALS AND RESULTS}
\subsection*{A. Datasets and Preprocessing}
\textbf{PD Datasets}:  The dataset from the University of New Mexico (UMN)\cite{pd}
, comprises EEG recordings of 27 PD patients and 27 healthy subjects. The dataset was acquired using 64-channel Ag/AgCl electrodes with the Brain Vision system at a sampling rate (frequency) of 500 Hz.

\textbf{MCI Dataset}: The dataset from a hospital in city A, includes 46 MCI patients and 43 healthy subjects. The dataset was acquired using 62 Ag/AgCl electrodes with the Brain Vision system at a sampling rate (frequency) of 5000 Hz.
\subsection*{B. Experiments Setup}
Our computational environment is set up with PyTorch 2.0.1 and Python 3.10.1, using an NVIDIA RTX 3090 for training. For the PD dataset, segments are defined with 500 sampling points, and we use 3-fold cross-validation. For the MCI dataset, segments are defined with 2000 sampling points each, and we employ 4-fold cross-validation for the experiments. For both datasets, we predict the label for each segment and then aggregate these predictions for the corresponding subject to assess the individual's disease status.

\subsection*{C. Experimental Results}
To validate the effectiveness of our experimental model, we compared it with various feature engineering methods, supervised and semi-supervised learning approaches, including correlation-based \cite{21} and SPD-based  \cite{22} functional connectivity features, Tensor-CSPNet \cite{23}, MAtt \cite{19}, COMET \cite{24}, CTW \cite{25}, and BNMTrans \cite{20}. We pre-trained our model using the N-1 Folds unlabeled training set and fine-tuned it with only a small amount of labeled data (10\% of labeled data in N-1 Folds). As shown in TABLE \ref{SOTA}, the results demonstrated that our proposed EEG-ReMinD achieved state-of-the-art (SOTA) performance,
thereby validating the effectiveness of our SSL-based two-stage methods from the perspective of Riemannian dynamics in decoding neurodegenerative EEG signals.

To test the robustness of the proposed EEG-ReMinD against corrupted data, we compared it with BNMTrans and COMET. In the experiments, the training data was intact while the test data was corrupted in three ways: Continuous Segment Corruption, where data was corrupted in random and continuous T/2 time segments; Channel Corruption, where the first and second channels were corrupted; and Non-Continuous Random Corruption, where 50\% of sampling points in both channels and time were randomly corrupted in each sample. 
For our model, during the pre-training phase, we first compute the Riemannian geometric states sequence for the intact EEG signals. Then, we implement data masking to the original EEG signals and train the framework using these masked EEG signals. The model then reconstructs the Riemannian geometric states from the masked data. 
Finally, we calculate the MSE loss by comparing the reconstructed Riemannian geometric states with the original states obtained from the complete EEG signals. 
Reconstructing the Riemannian geometric states sequence helps the model maintain robustness against noise or anomalies in the input data. By preserving the core structure of the data across multiple layers, the model can recover essential information even with poor data quality, enhancing its adaptability to incomplete data. 
TABLE \ref{corruption} have demonstrated that our framework exhibits strong resistance to corrupted data, maintaining robust performance even under data degradation. 


To evaluate the effectiveness of ReMinD's key components, we conducted ablation experiments in a top-down manner, including
modifying our Manifold-Euclidean integrated state reconstruction algorithm using either pure manifold or Euclidean transformer-based SSL (ManiSSL/EucliSSL),
removing the SSL pre-training stage (w/o SSL),
removing the 3D geometric positional encodings (w/o 3D-Pos),
and removing the spatiotemporal convolutional filters (w/o Filter).

The experiments were performed on PD and MCI datasets, as summarized in Table \ref{Ablation}. We demonstrate the superior performance of our two-stage learning approach, the innovative nature of our state reconstruction method, and the essential role of the incorporated positional and filtering modules.
\section*{ IV. Conclusion}
In this study, we propose a novel two-stage EEG decoding framework named \textit{EEG-ReMinD}, which is initially pre-trained using self-supervised internal state reconstruction that incorporates 3-D geometric position information and Riemannian dynamic analysis. Three experiments validated on two different neurodegenerative EEG datasets demonstrate the efficacy of \textit{EEG-ReMinD} in learning from limited labels and addressing data corruption. Our proposed method offers new insights into semi-supervised EEG decoding strategies.

\end{document}